\documentstyle[aps,prb,twocolumn,epsf,graphics,psfig]{revtex}

\begin{document}
%% The following two lines should be there when using 'twocolumn'.
\twocolumn[\hsize\textwidth\columnwidth\hsize\csname @twocolumnfalse\endcsname
\title{Theory of Magnetic Anisotropy in ${\bf\rm III}_{1-x}{\bf\rm Mn}_{x}{\bf\rm
V}$ Ferromagnets}
\author{M. Abolfath$^{1}$, T. Jungwirth$^{2,3}$, J. Brum$^{4}$
and A.H. MacDonald$^{2}$}
\address{$^{1}$Department of Physics and Astronomy,
University of Oklahoma, Norman, OK 73019-0225 \\}
\address{$^{2}$Department of Physics,Indiana University, Bloomington, Indiana 47405}
\address{$^{3}$Institute of Physics ASCR, Cukrovarnick\'a 10,
162 00 Praha 6, Czech Republic}
\address{$^{4}$Department of Physics, UNICAMP, Campinas, Brazil}
\date{\today}
\maketitle
\begin{abstract}
We present a theory of magnetic anisotropy in ${\rm III}_{1-x}{\rm Mn}_{x}{\rm
V}$ diluted magnetic semiconductors with carrier-induced ferromagnetism.  The
theory is based on four and six band envelope functions models for the valence
band holes and a mean-field treatment of their exchange interactions with ${\rm
Mn}^{++}$ ions.  We find that easy-axis reorientations can occur as a function
of temperature, carrier density $p$, and strain.  The magnetic anisotropy in
strain-free samples is predicted to have a $p^{5/3}$ hole-density dependence at
small $p$, a $p^{-1}$ dependence at large $p$, and remarkably large values at
intermediate densities. An explicit expression, valid at small $p$, is given
for the uniaxial contribution to the magnetic anisotropy due to unrelaxed
epitaxial growth lattice-matching strains. Results of our numerical simulations
are in agreement with magnetic anisotropy measurements on samples with
both compressive and tensile strains. We predict that decreasing the hole
density in current samples will lower the ferromagnetic transition temperature,
but will increase the magnetic anisotropy energy and the coercivity.
\end{abstract}


%% The following line should be there when using 'twocolumn'.
% Comment on six band models.
% Comment on beyond mean-field theory
% Discuss importance of disorder
% Check conversion from N_0\beta to J_{pd}
\vskip2pc]

\section{Introduction}

The discovery of carrier-mediated ferromagnetism\cite{ohno92,ohno96,review} in
${\rm III}_{1-x}{\rm Mn}_{x}{\rm V}$ and doped\cite{haury97} ${\rm
II}_{1-x}{\rm Mn}_{x}{\rm VI}$ diluted magnetic semiconductors (DMS's) has
opened up a broad and relatively unexplored frontier for both basic and applied
research. Experiments\cite{ohno96,munekata93} in ${\rm Ga}_{1-x}{\rm
Mn}_{x}{\rm As}$ and ${\rm In}_{1-x}{\rm Mn}_{x}{\rm As}$ have demonstrated
that these ferromagnets have remarkably square hysterisis loops with
coercivities typically $\sim 40 {\rm Oe}$, and that the magnetic easy axis is
dependent on epitaxial growth lattice-matching strains.  In this paper we
discuss the magnetic anisotropy properties of ${\rm III}_{1-x}{\rm Mn}_{x}{\rm
V}$ DMS ferromagnets, predicted by a mean-field-theory\cite{ourprb} of the
exchange interaction coupling between localized magnetic ions and valence band
free carriers.  We use phenomenological four or six band envelope function
models, depending on the carrier density $p$, in which the valence band holes
are characterized by Luttinger , spin-orbit splitting, and strain-energy
parameters.

The physical origin of the anisotropy energy in our model is spin-orbit
coupling in the valence band. Our work is based in part on theoretical
descriptions developed by Gaj\cite{gaj} and Bastard\cite{bastard} to explain the
optical properties of undoped, paramagnetic DMS's. As the critical temperature
is approached, the mean-field-theory we employ\cite{ourprb} reduces to  an
earlier theory\cite{haury97} that invokes generalized RKKY carrier-mediated
interactions between localized spins.  The two approaches differ, however, in
their description of the magnetically ordered state. As this work was nearing
completion we learned of a closely related study\cite{dietl00} which uses the
same mean-field theory to address critical temperature trends in this material
class and which also comments on the mean-field theory's ability to address
magnetic anisotropy physics. We are aware of three elements of the physics of
these materials which make the predictions of our mean-field theory uncertain:
i) we do not account for the substantial disorder which is usually present in
these ferromagnets; ii) we do not include effects due to interactions among the
itinerant holes and iii) we do not account for correlations between localized
spin-configurations and itinerant hole states.  The importance of each of these
deficiencies is difficult to judge in general, and probably depends on
adjustable material parameters.  In our view, it is likely that there is a
substantial range in the parameter space of interest where the predictions of
the present theory are useful.  We expect that important progress can be made
by comparing this simplest possible theory of carrier-induced DMS
ferromagnetism with experiment.

This work has two objectives.  Most importantly, we have attempted to shed
light on how various adjustable material parameters can influence magnetic
anisotropy.  Secondarily, we have made an effort to estimate the magnetic
anisotropy energy in those cases where experimental information is presently
available.  Our hope here is to initiate a process of careful and quantitative
comparison between mean-field theory and experiment, partially to help judge
the efficiency of this approximation in predicting other physical properties.
Even in the mean-field theory, we find that the magnetic anisotropy physics of
these materials is rich.  We predict easy axis reorientations as a function of
hole density, exchange interaction strength, temperature, and strain and
identify situations under which ${\rm III}_{1-x}{\rm Mn}_{x}{\rm V}$
ferromagnets are remarkably hard.

In Section II we detail our mean-field-theory of the ordered state.  The theory
simplifies in the limit of low-temperature and low hole densities. Our results
for this limit, presented in Section III, predict a $\langle111\rangle$ easy
axis in the absence of strain, and a magnetic anisotropy energy which is
approximately $10\%$ of the free carrier band-energy density.  This value is
extremely large for a cubic metallic ferromagnet; typical ratios in
transition metal ferromagnets are smaller than $10^{-6}$ for example.
The anisotropy energy in this limit varies as the free-carrier density to the
$5/3$ power and is independent of the exchange coupling strength.  Explicit
results for the strain-dependence of the magnetic anisotropy in the same limit
are presented in Sec. IV.  We find that unrelaxed lattice-matching strains due
to epitaxial growth contribute a uniaxial anisotropy which favors magnetization
orientation along the growth direction when the substrate lattice constant is
larger than the ferromagnetic semiconductor lattice constant and an in-plane
orientation in the opposite case.  
Unfortunately, perhaps, the simple low density limit does not normally
apply in situations where high critical temperatures are expected. The more
complicated, and more widely relevant, general case is discussed in Section V.
We find that magnetic anisotropy has a non-trivial dependence on both
temperature and exchange coupling strength and that easy axis reversals occur,
in general, as a function of either parameter. According to our theory,
anisotropy energy densities comparable to those in typical metallic
ferromagnets are possible when the exchange coupling is strong enough to
depopulate all but one of the spin-split valence bands, even with saturation
magnetization values smaller by more than an order of magnitude. In the limit
of large hole densities, we find that the anisotropy energy of strain-free
samples is proportional to hole density $p^{-1}$ and exchange coupling
$J_{pd}^4$. We find that in typical situations a strain $e_0$ of only $\sim 1\%$
is sufficient to overwhelm the cubic anisotropy of strain-free samples. We
conclude in Section VI with a discussion of the implications of these
calculations for the interpretation of present experiments, and with some
suggestions for future experiments which could further test the appropriateness
of the model used here.

\section{Formal Theory}

Our theory is based on an envelope function description of the valence band
electrons, and a spin representation for their kinetic-exchange
inteaction\cite{dmsreviews} with localized $d$ electrons\cite{commentonmodel}
on the Mn$^{++}$ ions:
\begin{equation}
{\cal H} = {\cal H}_m + {\cal H}_b + J_{pd} \sum_{i,I} {\vec S_I} \cdot {\vec
s}_i \; \delta({\vec r}_i - {\vec R}_I), 
\label{coupling}
\end{equation}
where $i$ labels a valence band hole and $I$ labels a magnetic ion. In
Eq.~(\ref{coupling}), ${\cal H}_m$ describes the coupling of magnetic ions with
total spin quantum number $J=5/2$ to an external field (if one is present),
$\vec S_I$ is a localized spin, $\vec s_i$ is a hole spin, and ${\cal H}_b$ is
either a four or a six-band envelope-function Hamiltonian\cite{Luttinger} for
the valence bands. In this paper we do not consider external magnetic fields so
${\cal H}_m \to 0$. The four-band Kohn-Luttinger model describes only the total
angular momentum $j=3/2$ bands, and is adequate when spin-orbit coupling is
large and the hole density $p$ is not too large. As discussed later, in the
case of GaAs, a four-band model suffices for $p \lesssim 10^{18} \;{\rm
cm}^{-3}$. In III$_{1-x}$Mn$_x$V semiconductors, the four $j=3/2$ bands are
separated by a spin-orbit splitting $\Delta_{so}$ from the two $j=1/2$ bands.
In the relevant range of hole and ${\rm Mn}^{++}$ densities, no more than four
bands are ever occupied. Nevertheless, mixing between $j=3/2$ and $j=1/2$ bands
does occur, and it can alter the balance of delicate cancellations which often
controls the net anisotropy energy. The exchange interaction between valence
band holes and localized moments is believed to be
antiferromagnetic\cite{dmsreviews}, i.e. $J_{pd}>0$. For GaAs,
experimental estimates\cite{ohno98,okabayashi98,matsukura98,omiya00} of
$J_{pd}$ fall between $0.04 {\rm eV}{\rm nm}^{3}$ and $0.15 {\rm eV}{\rm
nm}^{3}$, with more recent work suggesting a value toward the lower end of this
range.

The form of the valence band for Bloch wavevectors near the zone center in a
cubic semiconductor follows from ${\bf k}\cdot{\bf p}$ perturbation theory and
symmetry considerations.\cite{valencebands} The four band ($j=3/2$) and six
band ($j=3/2$ and 1/2) models are known as Kohn-Luttinger Hamiltonians and
their explicit form is given in the Appendix.  The eigenenergies are measured
down from the top of the valence band, i.e., they are hole energies. The
Kohn-Luttinger Hamiltonian contains the spin-orbit splitting parameter
$\Delta_{so}$ and three other phenomenological parameters, $\gamma_1$,
$\gamma_2$ and $\gamma_3$. These are accurately known for common
semiconductors. For GaAs and InAs, the two materials in which ${\rm
III}_{1-x}{\rm Mn}_{x}{\rm V}$ ferromagnetism has been observed,
$\Delta_{so}=0.34$~eV and 0.43~eV, and $(\gamma_1,\gamma_2,\gamma_3)=
(6.85,2.1,2.9)$ and $(19.67,8.37,9.29)$ respectively. Most of the specific
illustrative calculations discussed below are performed with GaAs parameters.

Our calculations are based on the Kohn-Luttinger Hamiltonian and on a
mean-field theory in which correlations between the local-moment configuration
and the itinerant carrier system are neglected.  We comment later on limits of
validity of this approximation.  There are a number of equivalent ways of
developing this mean-field theory formally.  In the following paragraphs we
present a view which is convenient for discussing magnetic anisotropy.

In the absence of an external magnetic field, the partition function of our
model may be expressed exactly as a weighted sum over magnetic impurity
configurations specified by a localized spin quantization axis, $\hat M$, and
azimuthal spin quantum numbers ${m_I}$:
\begin{equation}
Z = \sum_{m_I} \exp (-F_{b}[m_I]/k_B T), \label{partfunction}
\end{equation}
where $F_{b}[m_I]$ is the valence band free energy for holes which experience
an effective Zeeman magnetic field
\begin{equation}
{\vec h}(\vec r)[m_I] =  - J_{pd} \hat M \sum_{I} m_I \; \delta({\vec r} -
{\vec R}_I). \label{zeemanfield}
\end{equation}
The mean-field theory consists of replacing ${\vec h}(\vec r)[m_I]$ by its
spatial average for each magnetic impurity configuration, thereby neglecting
correlations between spin-distributions in local-moment and hole subsystems.
The effective Zeeman magnetic field experienced by the holes then depends only
on $\hat M$, the direction of the local-moment orientation, and the mean
azimuthal quantum number averaged over all local moments, $M$:
\begin{equation}
{\vec h}_{MF}(M) = J_{pd} N_{Mn} M \hat M \equiv h \hat M, \label{meanfieldh}
\end{equation}
where $N_{Mn} =N_I/V$ is the number of magnetic impurities per unit volume.
The mean-field partition function is
\begin{equation}
Z_{MF}(M) = \exp ( (N_I T s(M) - F_{b}(\vec h))/k_B T), \label{meanfieldz}
\end{equation}
where the entropy per impurity is defined by
\begin{equation}
s(M) = k_B \lim_{N_I \to \infty} \frac{\ln[ \sum_{m_I} \delta(\sum_I m_I - N_I
M)]}{N_I},
\end{equation}
and $F_{b}(\vec h)$ is the free-energy of a system of non-interacting fermions
with single-particle Hamiltonian $H_{b} - h \hat M \cdot \vec s$.

\begin{figure}
\center \epsfxsize 6.0cm \rotatebox{-90}{\epsffile{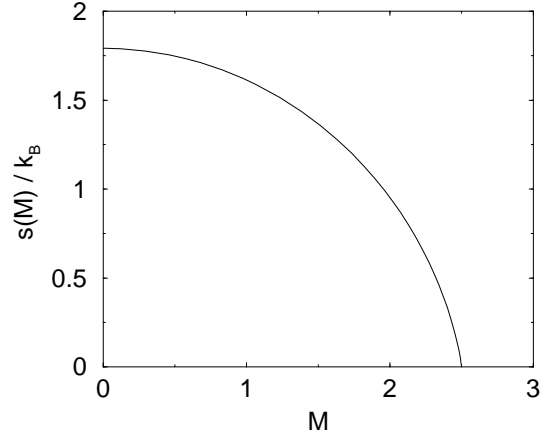}}
\vskip1cm \caption{Entropy per localized spin as a function of average
polarization. } \label{entropy}
\end{figure}

Following standard `large number' arguments $s(M)$ is readily evaluated by
considering an auxiliary system consisting of magnetic impurities coupled only
to an external magnetic field $H$.  For this model problem, a familiar
exercise\cite{ashcroftmermin} gives the result
\begin{equation}
M(H) = J B_J(x), \label{extfield}
\end{equation}
where $x = g_L \mu_B H J/ k_B T$, $g_L$ is the Land{\' e} g-factor of the ion,
$\mu_B$ is the electron Bohr magneton, and
\begin{equation}
B_J(x) = \frac{2J+1}{2J} \coth\big[(2J+1)x/2J\big] - \frac{1}{2J} \coth(x/2J),
\label{brillouin}
\end{equation}
is the Brillouin function.  The Brillouin function is a one-to-one mapping
between reduced fields $x$ in the interval $[0,\infty]$ and reduced
magnetizations $M/J$ in the interval $[0,1]$; the inverse function $B_J^{-1}$
maps $M/J$ to $x$. Since the magnetization maximizes $s(M) + g_L \mu_B H M /k_B
T$,
\begin{equation}
\frac{d s(M)}{d M} = - g_L \mu_B H/ k_B T. \label{dsdm}
\end{equation}
Eq.~(\ref{dsdm}) can be used to eliminate $H$ and arrive at the following
explicit expression for $s(M)$:
\begin{equation}
s(M) = k_B \int_{B_J^{-1}(M/J)}^{\infty} \; dx \; x \; \frac{dB_J(x)}{dx}.
\label{smexplicit}
\end{equation}
The $J=5/2$ result for $s(M)$ is illustrated in Fig.~\ref{entropy}. The entropy
per impurity vanishes for $M=J=5/2$ because there is a single configuration
with $\sum_I m_I = N_I J$, and approaches $\ln(2J+1) \approx 1.79$ for $M \to
0$.

%%%%%%%%%%%%%%%%%%%%%%%%%%%%%%%%%%%%%%%%%%%%%%%%%%%%%%%%%%%%%%
%%%%%%%%%%%%%%%%%%%%%%%%%%%%%%%%%%%%%%%%%%%%%%%%%%%%%%%%%%%%%%%%

The mean-polarization of the localized spins at a given temperature and for a
given orientation of the local moments is determined by minimizing the
mean-field free energy
\begin{eqnarray}
F_{MF}(M) &=& -k_B T \ln Z_{MF}(M) \nonumber \\
  &=& F_{b}(\vec h= N_{Mn} J_{pd} M \hat M) - k_B T N_I s(M),
\label{fmf}
\end{eqnarray}
with respect to $M$.  Setting the derivative to zero gives
\begin{equation}
\frac{d s(M)}{d M} = \frac{J_{pd}}{k_B T V} \frac{d F_{b}(h \hat M)}{d h}.
\label{minequation}
\end{equation}
Comparing with Eq.~(\ref{dsdm}), it follows that $F_{MF}(M)$ is minimized by
$M=JB_J(g_L \mu_B H_{eff} J/k_B T)= J B_J(x_{eff})$ where
\begin{equation}
x_{eff} \equiv \frac{g_L \mu_B H_{eff}}{k_B T} = - \frac{J_{pd}}{V k_B T}
\frac{d F_{b}(h \hat M)}{d h}. \label{heffequation}
\end{equation}
It follows that $h = N_{Mn} J_{pd} M$ is determined by solving the
self-consistent equation
\begin{equation}
h = N_{Mn} J_{pd} J B_J\big[x_{eff}(h)\big]. \label{meanfieldequation}
\end{equation}
Note that
\begin{equation}
\frac{d F_{b}(h \hat M)}{d h} = \langle \vec S_{tot} \cdot \hat M \rangle,
\label{hellmanfeynman}
\end{equation}
where $S_{tot}$ is the total hole spin and the angle brackets indicate a
thermal average for the non-interacting valence band system.  Since the valence
band system experiences an effective Zeeman coupling with strength proportional
to $h$ and direction $- \hat M$, it is clear that the right hand side of
Eq.~(\ref{hellmanfeynman}) is negative in sign and that its magnitude increases
monotonically with $h$, making it easy to solve Eq.~(\ref{meanfieldequation})
numerically.

To simplify the calculations presented in subsequent sections, we take
advantage of the fact that temperatures of interest are almost always
considerably smaller than the itinerant carrier Fermi energy.  This allows us
to replace $F_{b}(\vec h)$ by the ground state energy $E_{b}(\vec h)$.  Then,
using Eq.~(\ref{heffequation}) and Eq.~(\ref{fmf}), a single calculation of
$E_{b}(h \hat M)$ over the range from $h=0$ to $h = N_{Mn} J_{pd} J$ may be
used to determine the local-moment magnetization $M(T)$ and the free energy
$F(T) = F_{MF}(M(T))$ at {\em all} temperatures.

The mean-field theory critical temperature can be identified by linearizing the
self-consistent equation at small $h$.  We find that
\begin{equation}
k_B T_c(\hat M) = - \frac{J(J+1)}{3} \frac{N_{Mn} J_{pd}^2}{V} \frac{ d^2 F_{b}
(h \hat M)}{d h^2} \vert_{h=0}. \label{tcequation}
\end{equation}
The second derivative of the valence band free-energy with respect to field is
proportional to its Pauli spin-susceptibility, which is in turn proportional to
the valence band density of states at the Fermi energy, and to $p^{1/3}$ at
small $p$. In the absence of strain, it follows from cubic symmetry that the
right-hand-side of Eq.~(\ref{tcequation}) is independent of $\hat M$. Below the
critical temperature however, the mean-field free energy does depend on $\hat
M$; this dependence is the magnetic anisotropy energy we wish to calculate.  We
will see that the dependence of the anisotropy energy on hole density is very
different from that of the critical temperature.

\section{Magnetic Anisotropy in the strong exchange coupling limit}

Our mean-field theory simplifies at low temperatures and, for the four-band
model, simplifies further when $h$ is much larger than the characteristic
energy scale of occupied Kohn-Luttinger states.  A convenient typical energy
scale is the $h=0$ hole Fermi energy $\epsilon_{F0}$.  For a given value of
$N_{Mn} J_{pd}$ the largest value of $h$ is reached at $T=0$. Then, since
$H_{eff}$ is always non-zero, $x_{eff} \to \infty$ and the solution to the
mean-field equations is $M=J$, implying that $h=N_{Mn} J_{pd} J$ for every
orientation $\hat M$. At $T=0$ there is no entropic contribution to the
free-energy and
\begin{equation}
F_{MF}(T=0) = E_{b}(\vec h= N_{Mn} J_{pd} J\hat M). \label{fmfteq0}
\end{equation}
We note in passing that the magnetization density at $T=0$ has contributions
from the localized spins and the itinerant spins:
\begin{equation}
M_s(T=0) = 2 \mu_B J N_{Mn} + \frac{\kappa}{3V} \langle \vec S_{tot} \cdot \hat
M \rangle,
\end{equation}
where $\kappa$ is an additional parameter of the Luttinger model. Because of
the antiferromagnetic exchange interaction, the two terms here will tend to
have opposite signs with the first term, which is independent of the hole
density, typically very dominant. For $h \gg \epsilon_{F0}$ two further
simplifications occur. When the splitting of the hole bands by the effective
Zeeman coupling is sufficiently large, or the hole density $p$ is sufficiently
small, only the lowest energy hole band will be occupied.  Furthermore, the
effective Zeeman-term will dominate the mean-field single-particle Hamiltonian
and, as we detail below, the envelope function spinor for this occupied hole
state has a simple analytic expression.  In this section we assume that the
spin-orbit splitting energy $\Delta_{so}$ is much larger than all other
energies so that we can work with a four band model.  More generally, the
anisotropy will depend on $h/\Delta_{so}$, even in the limit of small hole
densities.

To judge whether or not this limit can be achieved in practice, we estimate the
Fermi energy of holes using the spherical approximation\cite{spherical} in
which the doubly degenerate $h=0$ bulk heavy hole and light hole bands are
parabolic with masses $m_h=m/(\gamma_1-2 \bar \gamma) \sim 0.498 m$ and
$m_l=m/(\gamma_1+2 \bar \gamma)  \sim 0.086 m$ respectively. An elementary
calculation then gives
\begin{equation}
\epsilon_{F0} = \frac{\hbar^2}{2m} \big( \frac{3 \pi^2 p}{2} \big)^{2/3} \bar
\gamma_0, \label{epsilonf0}
\end{equation}
where $n=N_h/V$ is the free-carrier density and
\begin{equation}
\bar \gamma_0 = \big[ \frac{(\gamma_1- 2 \bar \gamma)^{-3/2} + (\gamma_1+2 \bar
\gamma)^{-3/2}}{2} \big]^{-2/3}. \label{gamma0}
\end{equation}
For GaAs $\bar \gamma_0 \sim 3.05 $: the Fermi energy is the same as that of a
system with four identical effective mass $m/\gamma_0$ bands.  Typical high
$T_c$ ${\rm Ga}_{1-x}{\rm Mn}_x{\rm As}$ ferromagnetic semiconductor samples
have $p \sim 0.1 {\rm nm}^{-3}$ and $N_{Mn} \sim 1.0 {\rm nm}^{-3}$ ($x \sim
0.05$), although these parameters can presumably be varied widely.  Choosing a
$J_{pd}$ value in the mid-range of estimated values ($\sim 0.006 {\rm Ry}\;
{\rm nm}^3$) these parameters imply that $h \sim 0.015 {\rm Ry}$ and
$\epsilon_{F0} \sim 0.01 {\rm Ry}$. $h$ is neither large compared to
$\epsilon_{F0}$, nor small compared to $\Delta_{so}$. Thus, the simple
expressions discussed in this section are not accurate for current high $T_c$
systems. As our ability to engineer materials improves it should, however, be
possible to grow samples which {\em are} in the limit discussed here. Since $h$
is comparable to $\epsilon_{F0}$, we know, even before performing detailed
calculations, that valence band quasiparticle spectra in paramagnetic and
ferromagnetic states will differ qualitatively.

In the large $h$ limit the lone occupied spinor at each wavevector $\vec k$
will be the member of the $j=3/2$ manifold for which the total angular momentum
is aligned in the direction $ - \hat M$; the origin of the minus sign here is
the antiferromagnetic nature of the interaction between localized spins and
hole spins.  Explicit expressions for the expansion of such a spin coherent
states in terms of the eigenstates of $j_z$ are known\cite{auerbach}:
\begin{eqnarray}
\vert-\hat M\rangle  &=&  u^3|3/2\rangle + \sqrt{3} [u^2 v |1/2\rangle + u v^2
|-1/2\rangle] \nonumber \\ &+& v^3 |-3/2>. \label{spincoherentstate}
\end{eqnarray}
In Eq.~(\ref{spincoherentstate}) $u = - i \exp (-i \phi/2) \sin(\theta/2)$ and $v
= i \exp(i \phi/2) \cos(\theta/2)$ where $\theta$ and $\phi$ are the spherical
coordinates which specify the unit vector $ - \hat M$. In the large
$h/\epsilon_{F0}$ limit, the band term in the single-particle Hamiltonian may
be treated using first order perturbation theory.  Taking the expectation value
of the Kohn-Luttinger Hamiltonian in the spin coherent state we find that
\begin{eqnarray}
\epsilon(\vec k) &=& -\frac{h}{2} +  \langle -\hat M| H_{L}(\vec k) |- \hat
M\rangle \nonumber \\ &\equiv& -\frac{h}{2} + \frac{\hbar^2 k^2}{2m}
\gamma(\hat M, \hat k). \label{sceofk}
\end{eqnarray}
The first term on the right hand side of Eq.~(\ref{sceofk}) reflects the spin
coherent state property\cite{merzbacher}, $\hat M \cdot s |-\hat M\rangle  =
-|- \hat M \rangle/2$. In Eq.~(\ref{sceofk}) we have noted that for any $\hat M$
and any $\hat k$, the dependence of hole energy on $k=|\vec k|$ is quadratic.
Using this property, it follows that the Fermi energy
\begin{equation}
\epsilon_{F}(\hat M) = \frac{\hbar^2}{2 m} (6 \pi^2 p)^{2/3} \; \bar
\gamma(\hat M), \label{scef}
\end{equation}
and that the ferromagnetic ground state energy density is
\begin{equation}
\frac{E_{b}(\vec M)}{V} = -\frac{h p}{2} + \frac{3}{5} p \; \epsilon_{F}(\hat
M),
\end{equation}
where
\begin{equation}
\bar \gamma (\hat M) \equiv \big[ \int \frac{ d \hat k}{4 \pi} \; (\gamma(\hat
M, \hat k))^{-3/2} \big]^{-2/3}.
\end{equation}

In analogy with the $h=0$ quantity $\bar \gamma_0$ defined in Eq.~(\ref{gamma0}),
$ \bar \gamma (\hat M)$ is an average of the band energy curvature over
reciprocal space directions $\hat k$, with the smaller values of $\gamma(\hat
M,\hat k)$ weighted more heavily.  Note that the factor $3 \pi^2/2$ in
Eq.~(\ref{epsilonf0}) is replaced by $6 \pi^2$ in Eq.~(\ref{scef}) because only one
band is occupied in this limit, instead of the four which are occupied at
$h=0$.  $m/\bar \gamma(\hat M)$ may be thought of as a spin-orientation
dependent effective mass, which is readily evaluated as a function of $\hat M$,
given the Luttinger parameters of any material.  Although the magnetic
condensation energy has a term proportional to $J_{pd}$, only the band-energy
contributes to the $\hat M$ dependence of the ferromagnetic ground state
energy.  The magnetic anisotropy energy in this limit is independent of
$J_{pd}$ and proportional to the hole density $p$ to the 5/3 power.

We have evaluated $\gamma(\hat M)$ as a function of angle for the Luttinger
parameters of GaAs and InAs.   As discussed in more detail later, we always
find that magnetic anisotropy in the absence of strain is well described by a
cubic harmonic expansion truncated at sixth order, an approximation commonly
used in the literature\cite{skomskicoey} on magnetic materials.  The
corresponding cubic harmonic expansion for $\bar \gamma(\hat M)$ is
\begin{eqnarray}
\bar \gamma(\hat M) &=&\bar \gamma(\langle100\rangle) + \gamma^{ca}_1 ({\hat
M}_x^2  {\hat M}_y^2 + {\hat M}_y^2 {\hat M}_z^2 + {\hat M}_x^2 {\hat M}_z^2)
\nonumber \\ &+& \gamma^{ca}_2 \; {\hat M}_x^2 {\hat M}_y^2 {\hat M}_z^2.
\label{gammaofm}
\end{eqnarray}
In our calculations we find that $\bar \gamma(\hat M) < \gamma_1$ for all
directions $\hat M$.  This property reflects the fact that small curvature
(large Fermi wavevector) directions are weighted more highly in calculating the
total hole energy.  For both InAs and GaAs we find that the dominant fourth
order cubic anisotropy coefficient, $\gamma^{ca}_1 <0$, indicating {\em
nickel-type} anisotropy with easy axes along the $\langle111\rangle$ cube
diagonal directions.  At a qualitative level, the source of the higher total
hole energy when the moment orientation is along a $\langle001\rangle$ (cube
edge) direction is easy to understand.  With such a moment orientation, the
occupied hole orbital has $j_z=-3/2$ and energy dispersion given by
$H_{hh}(\vec k)$ in Eq.~(\ref{lutpar}).  It follows that $\gamma(\hat
M=\langle100\rangle,\hat k)$ has the relatively large value $\gamma_1+\gamma_2$
for all orientations of $\hat k$ in the $\hat x -\hat y$ plane.  These large
values of $\gamma(\hat M,\hat k)$ are important in the average and cause the
average over $\hat k$ to reach its maximum for this orientation of $\hat M$.

The cubic magnetic anisotropy energy coefficients in this limit are given by
\begin{equation}
K^{ca}_{i} =  + \frac{3}{5} p \frac{\hbar^2}{2 m} (6 \pi^2 p)^{2/3} \;
\gamma^{ca}_{i}. \label{anisocoefficients}
\end{equation}
Values of $\gamma^{ca}_{i}$ for GaAs and InAs are listed in Table I.  We will
see later that these expressions apply up to $p \sim 10^{18} {\rm cm}^{-3}$.
Inserting this value for the hole density gives coefficients $\sim 2 {\rm
kJ}{\rm m}^{-3}$ for GaAs host material and $\sim 4 {\rm kJ}{\rm m}^{-3}$ for
InAs host materials; magnetic anisotropy is twice as strong in InAs in the
strain free case.  These anisotropy energy coefficients are not so much smaller
than those of the cubic metallic transition metal ferromagnets, despite the
much higher carrier densities in the metallic case. We will see below that the
scale of the semiconductor magnetic anisotropy energy does not change as the
carrier density increases from $10^{18} {\rm cm}^{-3}$ to $\sim 10^{21} {\rm
cm}^{-3}$.  The relatively large anisotropy energies occur despite the fact
that the saturation moments $M_s$ of ${\rm III}_{1-x}{\rm Mn}_x{\rm V}$
ferromagnets are more than an order of magnitude smaller than their cubic metal
counterparts. It follows from these values that the magnetic hardness
parameters of the ${\rm III}_{1-x}{\rm Mn}_x{\rm V}$ ferromagnets,
\begin{equation}
\kappa \sim \big[\frac{K^{ca}_1}{\mu_0 M_s^2}\big]^{1/2}, \label{hardness}
\end{equation}
will typically be larger than one.  This is unusual in cubic materials and
occurs because spin-orbit coupling has a much stronger influence on
semiconductor valence bands than on transition metal $d$-bands.

As we discuss at length in Section V, magnetic anisotropy does not continue to
increase rapidly with hole density once two or more bands are occupied in the
metallic state. In current high $T_c$ samples, we will find that several bands
are always occupied, even at zero temperature. The simple limit discussed in
this section demonstrates that anisotropy energies, $T=0$ saturation moments,
and critical temperatures will have radically different dependencies on
engineerable parameters.

\section{Strain Dependence of Magnetic Anisotropy: Low Hole Density Limit}

Because of the low soluability of Mn in III-V semiconductors, ${\rm
III}_{1-x}{\rm Mn}_x{\rm V}$ materials with $x$ large enough to produce
cooperative magnetic effects cannot be obtained by equilibrium growth. The MBE
growth techniques which have been successfully developed\cite{ohno89} produce
${\rm III}_{1-x}{\rm Mn}_x{\rm V}$ films whose lattices are locked to those of
their substrates.  X-ray diffraction studies\cite{review} have established that
the resulting strains are not relaxed by dislocations or other defects, even
for thick films. Strains in the ${\rm III}_{1-x}{\rm Mn}_x{\rm V}$ film break
the cubic symmetry assumed in the previous section. Fortunately, the influence
of MBE growth lattice-matching strains on the hole bands of cubic
semiconductors is well\cite{valencebands,pikus} understood.  For the
$\langle001\rangle$ growth direction used to create ${\rm III}_{1-x}{\rm
Mn}_x{\rm V}$ films, strain generates a purely diagonal contribution to the
four band single-particle envelope function Hamiltonian in the representation
we use in this paper, adding contributions $\delta \epsilon_{h}$ and $\delta
\epsilon_{l}$ respectively to $j_z=\pm 3/2$ heavy hole and $j_z= \pm 1/2$ light
hole entries. The energy shifts are related to the lattice strains by
\cite{valencebands}
\begin{equation}
\delta \epsilon_{h} = \frac{e_0}{C_{11}}\big[ - 2 a_{1} (C_{11} -C_{12}) -
\frac{a_{2}}{2} (C_{11}+2 C_{12}) \big], \label{deltahh}
\end{equation}
\begin{equation}
\delta \epsilon_{l} = \frac{e_0}{C_{11}}\big[ - 2 a_{1} (C_{11} -C_{12}) +
\frac{a_{2}}{2} (C_{11}+2 C_{12}) \big], \label{deltalh}
\end{equation}
where $e_0$ is the in-plane strain produced by the substrate-film lattice
mismatch:
\begin{equation}
e_0 = \frac{a_{S}-a_{F}}{a_{F}}. \label{strain}
\end{equation}
In Eqs.~(\ref{deltahh}--\ref{strain}), the $C_{ij}$ are the
elastic constants of the unstrained ${\rm III}_{1-x}{\rm Mn}_x{\rm V}$ film,
which we will assume to be identical to those of the host III-V material,
$a_{S}$ is the lattice constant of the substrate on which the ${\rm
III}_{1-x}{\rm Mn}_x{\rm V}$ film is grown, $a_{F}$ is the unstrained lattice
constant of bulk ${\rm III}_{1-x}{\rm Mn}_x{\rm V}$, and $a_{1}$ and $a_{2}$
are phenomenological deformation potentials whose values for common III-V
semiconductors are known.
%%%%%%%%%%%%%%%%%%%%%%%%%%%%%%%%%%%%%%%%%%%%%%%%%%%%%%%%%%%%%%%%%%%%%%%%
For the six band model the strain Hamiltonian includes off-diagonal elements
and is given up to a constant times the unit matrix by
\begin{equation}
H_{\rm strain} = \Gamma e_0\left(\begin{array}{cccccc} 0 & 0 & 0 & 0 & 0 & 0\\
0 & 1 & 0 & 0 & 0 & \frac{1}{\sqrt{2}}\\ 0 & 0 & 1 & 0 & -\frac{1}{\sqrt{2}} &
0 \\ 0 & 0 & 0 & 0 & 0 & 0\\ 0 & 0 & -\frac{1}{\sqrt{2}} & 0 & \frac{1}{2} &
0\\ 0 & \frac{1}{\sqrt{2}} & 0 & 0 & 0 & \frac{1}{2}\\
\end{array}\right),
\label{strainHamiltonian}
\end{equation}
where $\Gamma= \epsilon_{l} - \epsilon_{h} = a_2e_0(C_{11}+2C_{12})/C_{11}$.
For GaAs and InAs, $\Gamma=-0.2382$~Ry and $-0.2762$~Ry
respectively.\cite{valencebands}
%%%%%%%%%%%%%%%%%%%%%%%%%%%%%%%%%%%%%%%%%%%%%%%%%%%%%%%%%%%%%%%%%%%%%%%%

As in the previous section, we can derive an explicit expression for the strain
contribution to the magnetic anisotropy energy when $h \gg \epsilon_{F0}$ and
$h \gg \Gamma$, allowing band and strain terms to be treated as a perturbative
correction to the effective Zeeman coupling, and $h \ll \Delta_{so}$, allowing
a four-band model to be used. In this way we obtain
\begin{equation}
\frac{E_{strain}(\hat M)}{V} = \frac{p}{4} [3 \delta \epsilon_{h} + \delta
\epsilon_{l}]
 + \frac{3 p \cos^2 \theta}{4} [\delta \epsilon_{h} - \delta \epsilon_{l}].
\label{strainanisotropy}
\end{equation}
Strain produces a uniaxial contribution to the magnetic anisotropy of ${\rm
III}_{1-x}{\rm Mn}_x{\rm V}$ films, that favors orientations along the growth
direction when strain shifts heavy holes down relative to the light holes and
orientations in the plane in the opposite circumstance.  Using
Eq.~(\ref{deltahh}) and Eq.~(\ref{deltalh}) and deformation potential elastic
constant values\cite{valencebands}, we find that a contribution to the energy
density given by $K_{strain} \sin^2 (\theta)$ where the uniaxial anisotropy
constant is $K_{strain} = -0.36 {\rm Ry} \;e_0 \; p$ for GaAs and $K_{strain} =
-0.41 {\rm Ry} \; e_0\; p$ for InAs.  Since $a_2 < 0$, compressive strain ($e_0
< 0$) lowers the heavy hole energy relative to the light hole energy and favors
moment orientations in the growth direction while tensile strain ($e_0 > 0$)
favors moment orientations perpendicular to the growth direction.  Since the
lattice constant of ${\rm Ga}_{1-x}{\rm Mn}_xAs$ is larger\cite{ohno96} than
that of GaAs, while that of ${\rm In}_{1-x}{\rm Mn}_xAs$ is smaller than that
of InAs, ${\rm Ga}_{1-x}{\rm Mn}_xAs$ on GaAs is under compressive strain and
${\rm In}_{1-x}{\rm Mn}_xAs$ on InAs is under tensile strain.
Assumming\cite{review} Vergard's law, $e_0 = .0004$ and $e_0=-.0028$ for InAs
and GaAs respectively at $x=0.05$.  For $p = 10^{-3} {\rm nm}^{-3}$, a density
for which this low-density result is still reasonably reliable, we find that
$K_{strain} = -0.36 {\rm kJ} {\rm m}^{-3}$ for InAs and $K_{strain} = 2.2 {\rm
kJ} {\rm m}^{-3}$ for GaAs.  At these densities, strain and cubic band
contributions are comparable in GaAs, but the latter contribution is dominant
in the InAs case.

Strain-induced anisotropy energies can be modified by growing the magnetic film
on relaxed buffer layers.  This effect has been demonstrated\cite{ohnosemic} by
Ohno {\em et al.} who showed that the easy axis for ${\rm Ga}_{1-x}{\rm
Mn}_xAs$ films changes from in-plane to growth-direction when the substrate is
changed from GaAs to relaxed (In,Ga)As buffer layers.   For $15\%$ In, the
magnetic film strain changes from compressive $e_0=-0.0028$ to tensile
$e_0=0.0077$ when this change is made.  We note that the sense of this change
is opposite to that predicted by our large $h$, small $p$ analytic result which
predicts that compressive strains favor growth direction orientations.
Similarly, ${\rm In}_{1-x}{\rm Mn}_xAs$ on InAs is under a small tensile strain
but is observed to have a growth direction easy axis. (Recall that in the large
$h$ limit the band anisotropy makes the growth direction the hard axis.)  This
distinction may be taken as an experimental proof that several hole bands are
occupied in the magnetic ground state of these materials. The strain anisotropy
energy {\em does} change sign at smaller values of $h$ partly because the first
band to be depopulated has primarily heavy-hole character. We will see in the
next section that the mean-field does predict the correct sign for the strain
anisotropy energy at experimental hole densities. We conclude from the present
considerations that strain can play a strong role in band-structure-engineering
of ${\rm III}_{1-x}{\rm Mn}_x{\rm V}$ ferromagnet magnetic properties.

\section{Partially Polarized Hole Band States: Magnetic Anisotropy in the General Case}

The results for $F_{b}(h\hat M)$ discussed in the previous section become
accurate when the effective Zeeman coupling $h$ is large enough to reduce the
number of occupied hole bands to one, {\em and} the Fermi energy remains safely
smaller than the spin-orbit splitting.  The situation is much more complicated
at smaller $h$ and larger hole densities.   Then several bands are occupied,
even at $T=0$, and these usually give competing contributions to the magnetic
anisotropy. It is {\em not} true in general that band and strain contributions
to the magnetic anisotropy are simply additive. We expect that it will
eventually be possible to realize a broad range of material parameters, and hence
a broad range of $h$ values, in ${\rm III}_{1-x}{\rm Mn}_x{\rm V}$
ferromagnets. The range of possibilities is immense and accurate modeling of a
particular sample will require accurate values for $p$, $J_{pd}$, and $N_{Mn}$
for that material.

\begin{figure}
\center \epsfxsize 6.0cm \rotatebox{-90}{\epsffile{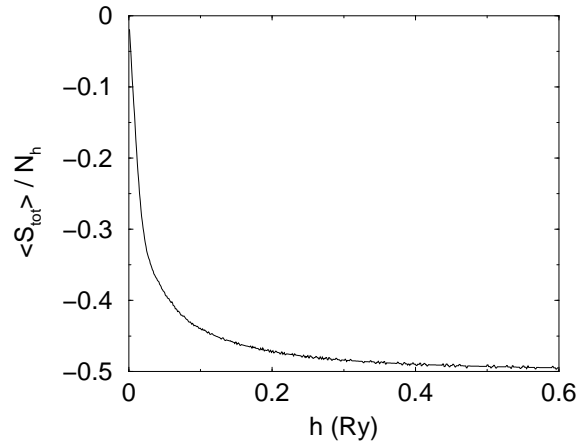}}
\vskip1cm \caption{Valence band spin per hole as a function of effective Zeeman
field strength $h$.} \label{holespinpolarization}
\end{figure}

In this section we discuss a series of illustrative calculations, starting with
ones performed using a four-band model of strain-free ${\rm Ga}_{1-x}{\rm
Mn}_x{\rm As}$ at hole density $p = 0.1 {\rm nm}^{-3}$. The valence band energy
density is
\begin{equation}
\frac{E_{b}}{V} = \frac{1}{V} \sum_{\bf k} \sum_{j=1}^{N_b} \theta(\epsilon_F -
\epsilon_{j}({\bf k}, \vec{h})) \epsilon_{j}({\bf k}, \vec{h}),
\label{holeenergy}
\end{equation}
where $\epsilon_F$ is the Fermi energy, $\vec k$ is the Bloch wavevector, and
the mean-field-theory quasiparticle energies $\epsilon_{j}({\bf k},\vec{h})$
are eigenvalues of the $N_b \times N_b$ single-particle Hamiltonian ($N_b$ is
the number of bands included in the envelope function Hamiltonian)
\begin{equation}
H_b = H_L + \vec{h} \cdot \vec{s} + H_{\rm strain}. \label{xxx1}
\end{equation}
For a strain free model we set $H_{\rm strain}=0$. In
Fig.~\ref{holespinpolarization} we plot the calculated spin-polarization per
hole as a function of $h$ for a growth direction field orientation; recall that
this quantity can be obtained by differentiating the energy per hole with
respect to $h$ and that the effective field seen by the localized spins is
obtained by multiplying this quantity by $J_{pd} \; p$.  The hole spin
polarization increases linearly at small $h$ with a slope proportional to the
valence band Pauli susceptibility.  We see that for the hole density of
Fig.~\ref{holespinpolarization}, complete hole spin-polarization is approached
only at values of $h$ comparable to or larger than the spin-orbit splitting of
GaAs, so that the four band model is not physical in this regime. For any given
model, and a given moment orientation, a {\em single} calculation of this type
provides all the microscopic information required to solve the mean-field
equations at all temperatures.
%%%%%%%%%%%%%%%% hole spin polarization plot &&&&&&&&&&&&&&&&&&&&&&&&&&&&&&&&&&
%%%%%%%%%%%%%%%%%%%%%%%%%%%%%%%%%%%%%%%%%%%%%%%%%%%%%%%%%%%%%%%%%%%%%%%%%%%%%%%

\begin{figure}
\center \epsfxsize 6.0cm \rotatebox{-90}{\epsffile{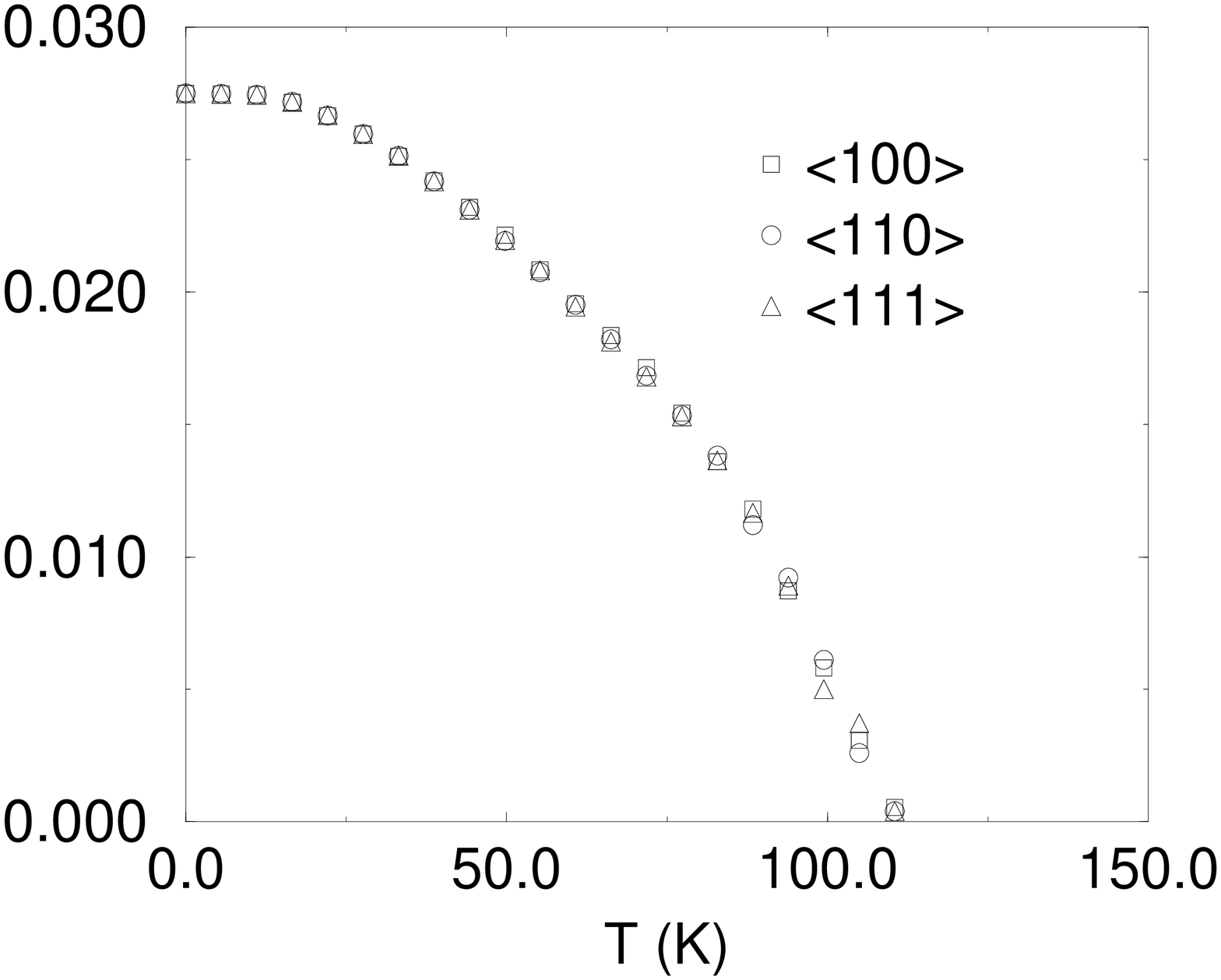}}
\vskip1cm \caption{ The dependence of the effective Zeeman coupling strength
seen by the magnetic impurities $h$ on temperature for the high symmetry
directions $<100>$,  $<110>$, and $<111>$.  These results were calculated using
a strain free four-band model for $J_{pd}=0.15 {\rm eV} {\rm nm}^{3}$ and
$N_{Mn}=1.0 {\rm nm}^{-3}$.  $h(T)$ is determined by solving the
self-consistent mean-field equation. } \label{Fig8}
\end{figure}

For a fixed values of $J_{pd}$, $N_{Mn}$, and $p$, $h$ must be evaluated as a
function of temperature by solving the self-consistent mean-field equation,
Eq.~(\ref{meanfieldequation}), and using numerical results like those plotted in
Fig.~\ref{holespinpolarization}. Results for $h(T)$ calculated for $J_{pd}=0.15
{\rm eV} {\rm nm}^{3}$, $N_{Mn}=1 {\rm nm}^{-3}$, and $p=0.1 {\rm nm}^{-3}$ are
illustrated in Fig.~\ref{Fig8} for high symmetry moment orientation directions.
For these parameters the critical temperature $T_c$ ($h(T) = 0$ for $T > T_c$)
is $\sim 100 {\rm K}$, in rough agreement with experiment. There are, however,
other choices of parameters which are also consistent with the measured
critical temperature.  In addition, as we discuss further in the next section,
it is not clear that this level of theory should always yield accurate results
for $T_c$.
%%%%%%%%%%%%%%%%%%%%%%%%%%%%%%%%%%%%%%%%%%%%%%%%%%%%%%%%%%%%%%%%%%%%%

\begin{figure}
\center \epsfxsize 6.0cm \rotatebox{-90}{\epsffile{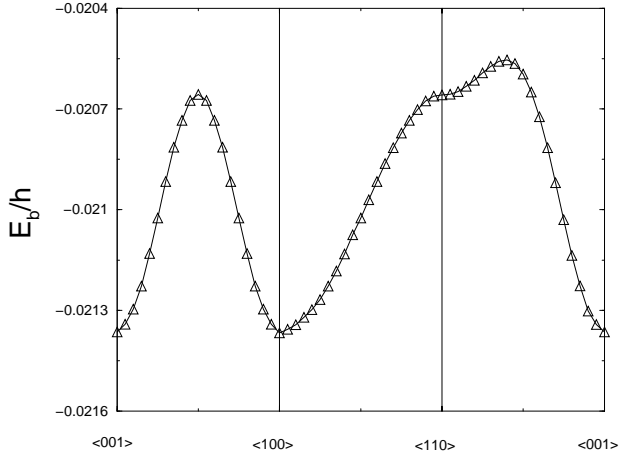}}
\vskip1cm \caption{Hole energy per particle in units of $h$ as a function of
magnetization orientation for $h=0.0275 {\rm Ry})$.  The solid line is a cubic
harmonic expansion fit to these results truncated at 6th order.  At this value of
$h$, three hole bands are partly occupied and the easy axes are the cube edge
directions. } \label{Fig5}
\end{figure}

\begin{figure}
\epsfxsize 8.4cm {\epsffile{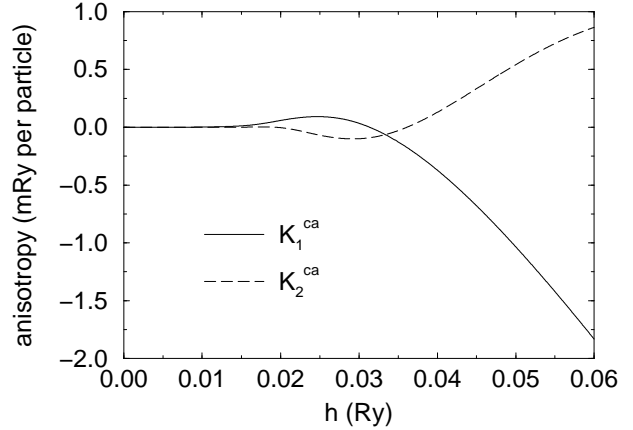}}
\vskip1cm \caption{ The dependence of the crystalline anisotropy coefficients,
$K_1^{ca}(h)$ and $K_2^{ca}(h)$ on $h$ for a four-band model with $p=0.1
{nm}^{-3}$. } \label{Fig7}
\end{figure}

At $T=0$, $h$ reaches its maximum value, $J_{pd} N_{Mn} J \sim 0.0275 {\rm
Ry}$. The dependence of energy on magnetization orientation at this value of
$h$ is illustrated in Fig.~\ref{Fig5} and compared with the cubic harmonic
expansion truncated at 6th order.  The coefficients of this expansion are fixed by
energy per volume calculations in $\langle100\rangle$, $\langle110\rangle$, and
$\langle111\rangle$ directions. In Fig.~\ref{Fig5}, and in all other cases we
have checked, the truncated cubic harmonic expansion is very accurate.  It is
therefore sufficient to evaluate the energy per volume in the high-symmetry
directions and to use
\begin{eqnarray}
K_{1}^{ca} &=& \frac{4 (E_{b}\langle110\rangle - E_{b}\langle100\rangle)}{V}
\nonumber \\ K_{2}^{ca} &=& \frac{27 E_{b}\langle111\rangle - 36
E_{b}\langle110\rangle + 9 E_{b} \langle100\rangle}{V}. \label{cubicanisotropy}
\end{eqnarray}
Results for $K_{i}^{ca}(h)$ obtained for the four band model in this way are
summarized in Fig.~\ref{Fig7}.  These results can be combined with those in
Fig.~\ref{Fig8} to obtain the model's cubic anisotropy coefficients as a
function of temperature, hole density and $h$. Here we see explicitly the
anisotropy reversals which commonly accompany hole band depopulations.  We note
in Fig.~\ref{Fig7} that the analytic results of Section III are recovered only
for very large values of $h$ at this density.

%%%%%%%%%%%%%%%%%%%%%%%%%%%%%%%%%%%%%%%%%%%%%%%%%%%%%%%%%%%%%%%%%%%%%
%%%%%%%%%%%%%%%%%%%%%%%%%%%%%%%%%%%%%%%%%%%%%%%%%%%%%%%%%%%%%%%%%%%%%%%%%%%%%
%%%%%%%%%%%%%%%%%%%%%%%%%%%%%%%%%%%%%%%%%%%%%%%%%%%%%%%%%%%%%%%%%%%%%%%%%%%%555

\begin{figure}
\center
\epsfxsize 6.8cm {\epsffile{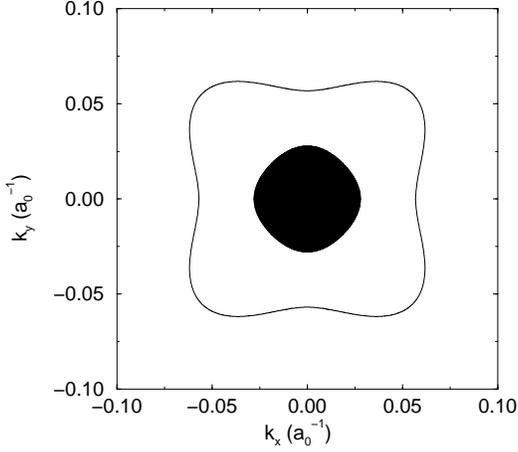}}
\vskip1cm \caption{Fermi surface
intersection with the $k_z=0$ plane for a four-band model with
$p=0.1 {\rm nm}^{-3}$ and $h=0$.  Doubly
degenerate heavy-hole (larger contour) and light-hole (smaller contour) bands
are occupied. } \label{fl_4_0}
\end{figure}

\begin{figure}
\center
\epsfxsize 6.8cm {\epsffile{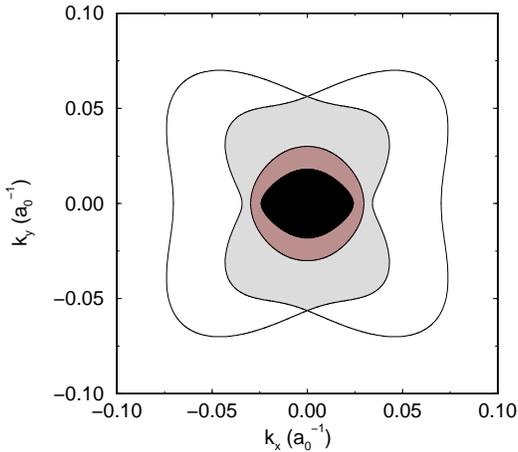}}
\vskip1cm \caption{Intersection of
the Fermi surface and the $k_z=0$ plane for a four-band model with
$p=0.1 {\rm nm}^{-3}$ and $h=0.01
{\rm Ry}$. This value of $h$ solves the mean-field equations equation at $T =0
K$, $N_{Mn}=1 {\rm nm}^{-3}$, and $J_{pd}=0.05 {\rm eV}{\rm nm}^3$, which is
near the lower experimental estimate \protect\cite{omiya00} for the exchange
coupling constant. (For $J_{pd}=0.15 {\rm eV}{\rm nm}^3$ this value of $h$
solves the mean-field equations at $T =85 K$). The magnetization orientation is
along the $\langle100\rangle$ direction. Heavy-hole and light-hole bands are
split at non-zero $h$. } \label{fl_4_0.01}
\end{figure}

\begin{figure}
\center
\epsfxsize 6.8cm {\epsffile{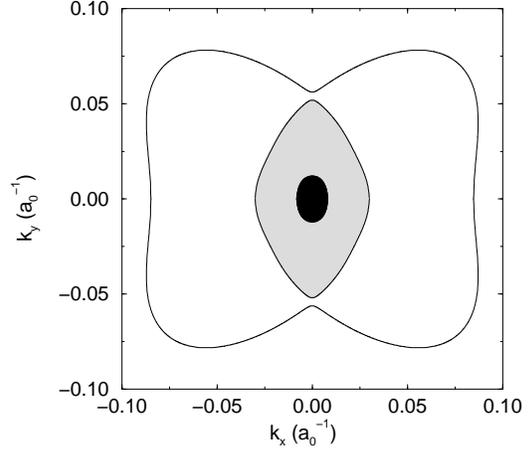}} \vskip1cm
\caption{Intersection of the Fermi surface and the $k_z=0$ plane for a
four-band model with $p=0.1 {\rm nm}^{-3}$ and $h=0.0275 {\rm Ry}$. This value
of $h$ solves the mean-field equation at $T =0 $, $N_{Mn}=1 {\rm nm}^{-3}$, and
$J_{pd}=0.15 {\rm eV}{\rm nm}^3$, which is  the upper experimental estimate
\protect\cite{omiya00} for the exchange coupling constant.  The magnetization
orientation is along the $\langle100\rangle$ direction. The minority heavy-hole
band is empty at this value of $h$. } \label{fl_4_0.0275}
\end{figure}

The valence band Fermi surfaces of ${\rm III}_{1-x}{\rm Mn}_x{\rm V}$
ferromagnets will be strongly dependent on both temperature and moment
direction orientation. Four-band model Fermi surface intersections with the
$k_z=0$ plane are illustrated in Figs.~\ref{fl_4_0}--\ref{fl_4_0.0275} for
$p=0.1 {\rm nm}^{-3}$ at $h=0$, $h=0.01 {\rm Ry}$, and $h=0.0275 {\rm Ry}$
respectively.
All figures are for moments oriented in the $\langle 100\rangle$ direction. 
The largest of
these three values of $h$ is the $T=0$ effective field ($J_{pd} N_{Mn} J$) for
$x=0.05$ and $J_{pd}$ at the high end of literature estimates. ($J_{pd}= 0.15
{\rm eV} {\rm nm}^3$.) These three figures represent the mean-field-theory
Fermi surfaces at three different temperatures.  In the spherical
approximation, the $h=0$ Fermi energy at this hole density is $\epsilon_{F0}
\sim 0.01 {\rm Ry}$, so a strong distortion of the bands is expected in the
magnetic state. At $h=0$, the hole bands occur in degenerate pairs; we refer to
the two less dispersive bands which occupy the larger area in $\hat k$ space as
heavy-hole bands, although this terminology lacks precise meaning in the
general case.  As $h$ increases, both heavy and light hole bands split.  For
small $h$, the minority-spin heavy-hole band occupation decreases rapidly and
all other band occupations increase. The heavy-hole minority spin band is
completely depopulated for $h \sim 0.04 {\rm Ry}$.  Once this band is empty,
the light-hole minority-spin Fermi radii begin to shrink rapidly. As seen in
Fig.~\ref{fl_4_0.0275}, this second band is nearly depopulated at $h=0.0275
{\rm Ry}$. At still stronger fields, the majority-spin light-hole band is
depopulated and the single-band limit addressed in preceding sections is
achieved. For $p=0.1 {\rm nm}^{-3}$, the single-band limit is achieved only at
$h$ values for which the four-band model breaks down.

%%%%%%%%%%%%%%%%%%%%%%%%%%%%%%%%%%%%%%%%%%%%%%%%%%%%%%%%%%%%%%%%%%%%%

%%%%%%%%%%%%%%%%%%%%%%%%%%%%%%%%%%%%%%%%%%%%%%%%%%%%%%%%%%%%%%%%%%%%%

\begin{figure}
\center
\epsfxsize 8.4cm {\epsffile{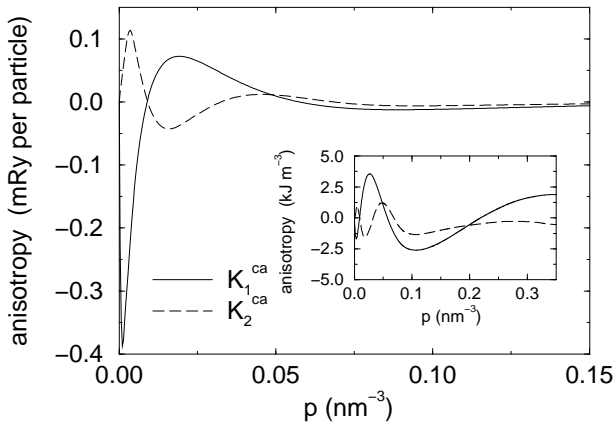}}
\vskip1cm \caption{Six-band model, $h=0.01 {\rm Ry}$, results for the cubic
magnetic anisotropy coefficients in units of Rydberg per particle (main plot)
and in kJ~per~m$^3$ (inset). } \label{aniso_6}
\end{figure}

\begin{figure}
\center
\epsfxsize 8.4cm {\epsffile{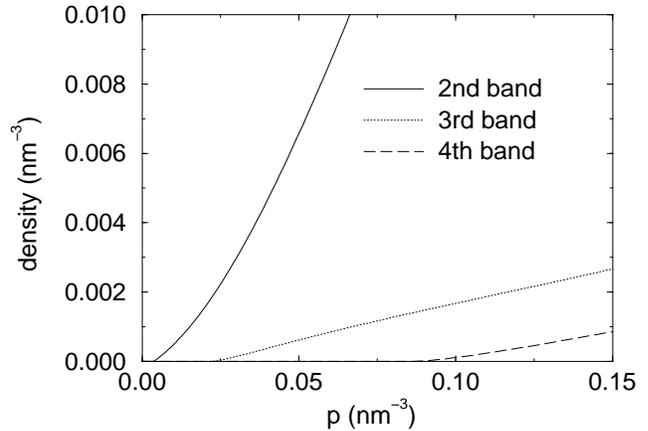}}
\vskip1cm \caption{Six-band model band densities as a function of the
total hole density; $h=0.01 {\rm Ry}$.}
\label{dennostrain}
\end{figure}

Finally we turn to a series of illustrative calculations intended to closely
model the ground state of ${\rm Ga}_{.95}{\rm Mn}_{.05}{\rm As}$. For this Mn
density and the smaller values of $J_{pd}$ favored by recent estimates, $h(T=0)
= J_{pd} N_{Mn} J \sim 0.01 {\rm Ry}$.  This value of $h$ is not so much
smaller than the spin-orbit splitting parameter in GaAs
($\Delta_{so}=0.025$~Ry), so that accurate calculations require a six band
model.   Even with $x$ fixed, our calculations show that the magnetic
anisotropy of ${\rm Ga}_{.95}{\rm Mn}_{.05}{\rm As}$ ferromagnets is strongly
dependent on both hole density and strain. The hole density can be varied by
changing growth conditions or by adding other dopants to the material, and
strain in a ${\rm Ga}_{.95}{\rm Mn}_{.05}{\rm As}$ film can be altered by changing
substrates as discussed previously. The cubic anisotropy coefficients (in units
of energy per volume) for strain-free material are plotted as a function of
hole density in the inset of Fig.~\ref{aniso_6}; the main plot shows the
coefficients in units of energy per particle. Over the density range $p < 0.05
{\rm nm}^{-3}$, four and six band models are in good agreement.  We see from
this result that the asymptotic low density region where the anisotropy energy
varies as $p^{5/3}$ holds only for $p < 0.005 {\rm nm}^{-3}$ at this value of
$h$.  The easy axis is nearly always determined by the leading cubic anisotropy
coefficient $K_1^{ca}$, except near values of $p$ where this coefficient
vanishes. As a consequence the easy-axis in strain free samples is almost
always either along one of the cube edge directions ($K_1^{ca} > 0$), or along
one of the cube diagonal directions ($K_1^{ca} < 0$). Transitions in which the
easy axis moves between these two directions occur twice over the range of hole
densities studied.  (Similar transitions occur as a function of $h$, and
therefore temperature, for fixed hole density.) Near the hole density
0.01~nm$^{-3}$, both anisotropy coefficients vanish and a fine-tuned isotropy
is achieved. The slopes of the anisotropy coefficient curves vary
as the number of occupied bands increases from $1$ to $4$ with
increasing hole density. This behavior is clearly seen from the correlation
between
oscillations of the anisotropy coefficients 
and onsets of higher band occupations, 
plotted in Fig.~\ref{dennostrain}.

\begin{figure}
\center
\epsfxsize 6.8cm {\epsffile{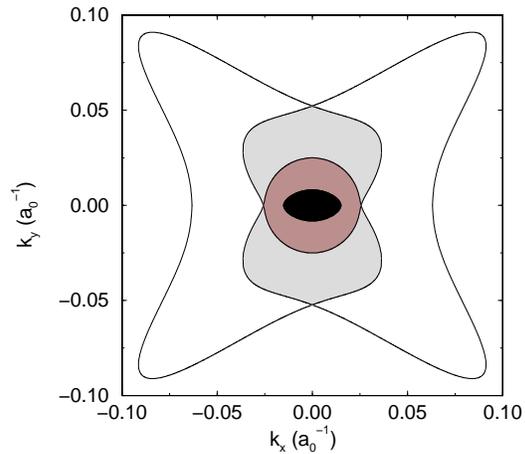}}
\vskip1cm \caption{Six-band model
Fermi surface intersections with the $k_z=0$ plane for $p=0.1 {\rm nm}^{-3}$
and $h=0.01 {\rm Ry}$.  This figure is for magnetization orientation is along
the $\langle100\rangle$ direction. } \label{fl_6_100}
\end{figure}

\begin{figure}
\center
\epsfxsize 6.8cm {\epsffile{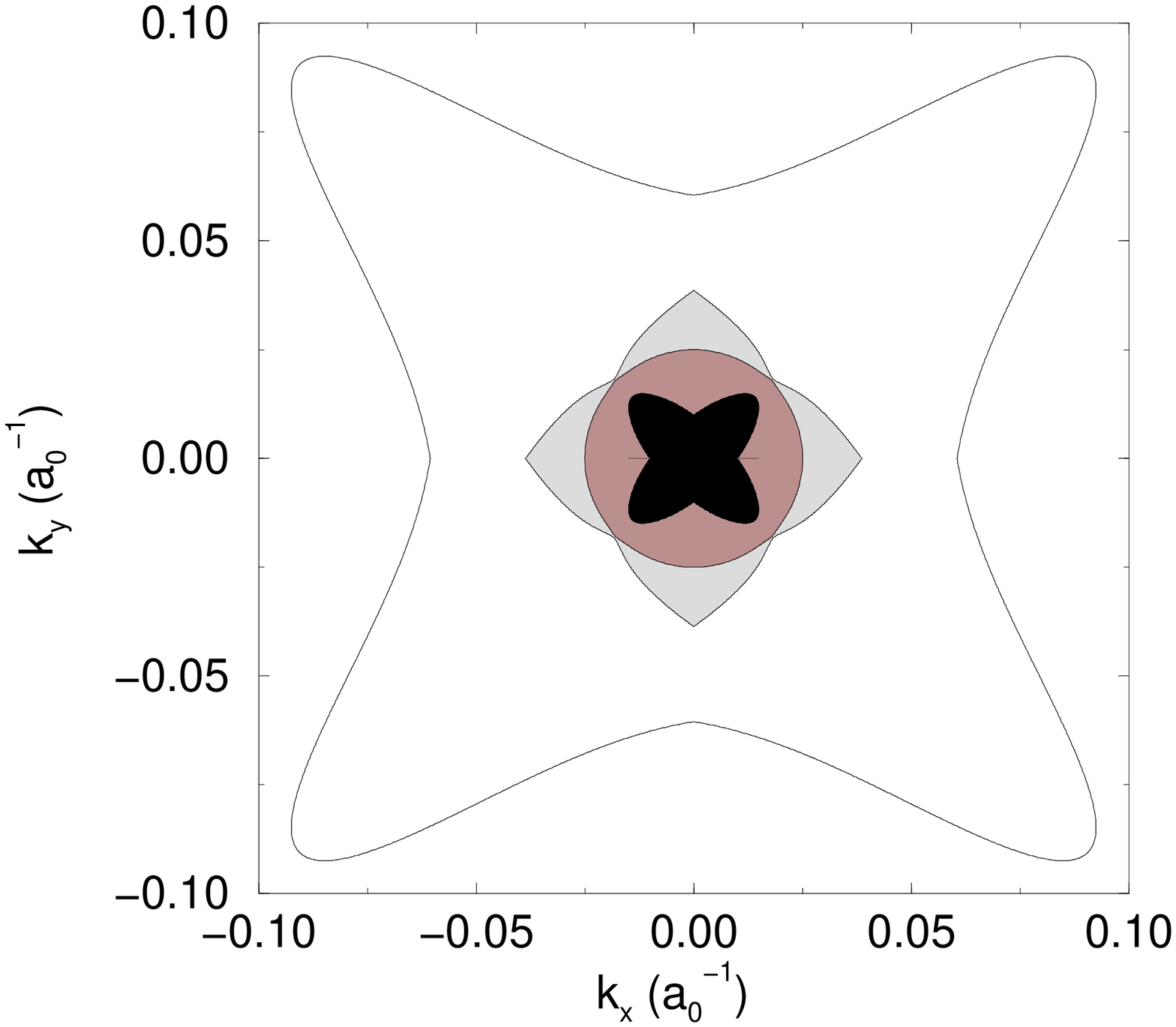}}
\vskip1cm \caption{Six-band model
Fermi surface intersections with the $k_z=0$ plane for the parameters of
figure~\protect\ref{fl_6_100} and magnetization orientation along the
$\langle110\rangle$ direction. } \label{fl_6_110}
\end{figure}

\begin{figure}
\center
\epsfxsize 6.8cm {\epsffile{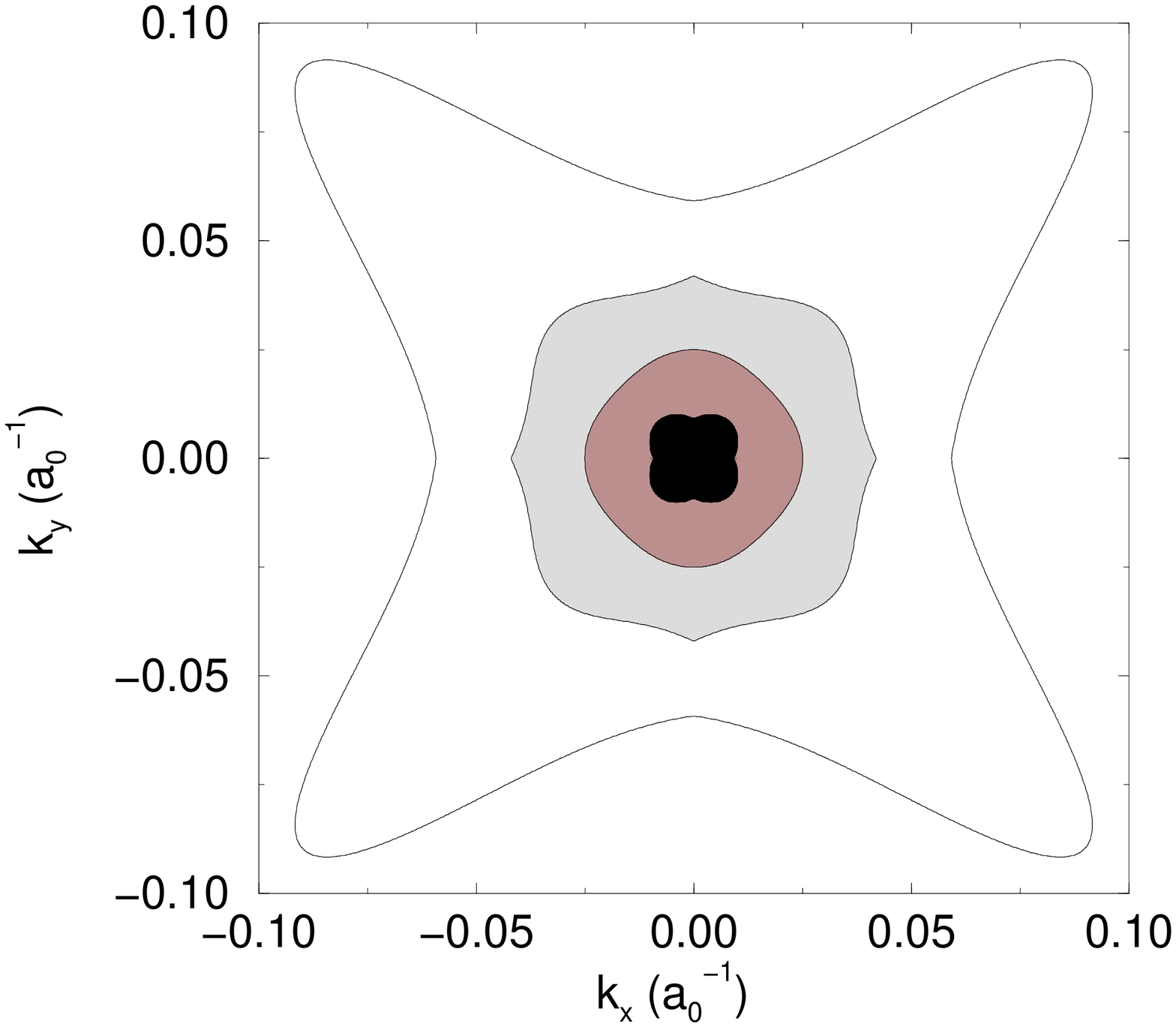}}
\vskip1cm \caption{Fermi surface
intersections with the $k_z=0$ plane for the parameters of
figure~\protect\ref{fl_6_100} and magnetization orientation along the
$\langle111\rangle$ direction. } \label{fl_6_111}
\end{figure}

Six-band model Fermi surfaces are illustrated in
Figs.~\ref{fl_6_100}-\ref{fl_6_111} by plotting their intersections with the
$k_z=0$ plane at $p =0.1 {\rm nm}^{-3}$ for the cases of $\langle100\rangle$,
$\langle110\rangle$, and $\langle111\rangle$ ordered moment orientations.
Comparing Fig.~\ref{fl_4_0.01} and Fig.~\ref{fl_6_100}, which differ only in
the band model employed, we see that there is a marked difference between the
majority-spin heavy hole bands in four and six-band cases.  For the six band
model, quasiparticle dispersion is particularly slow, leading to large Fermi
radii along the $\langle110\rangle$ directions.  The large and more anisotropic
mass is a consequence of mixing with the split-off hole bands.  As we see from
Figs.~\ref{fl_6_110} and \ref{fl_6_111}, this effect occurs for all ordered
moment orientations, although the details of the small minority band Fermi
surface projections change markedly. The dependence of quasiparticle band
structure on ordered moment orientation, apparent in comparing these figures,
should lead to large anisotropic magnetoresistance effects in ${\rm
III}_{1-x}{\rm Mn}_x{\rm V}$ ferromagnets.  We also note that in the case of
cube edge orientations, the Fermi surfaces of different bands intersect.  This
property could have important implications for the decay of long-wavelength
collective modes \cite{spin-wave}.


\begin{figure}
\center
\epsfxsize 8.4cm {\epsffile{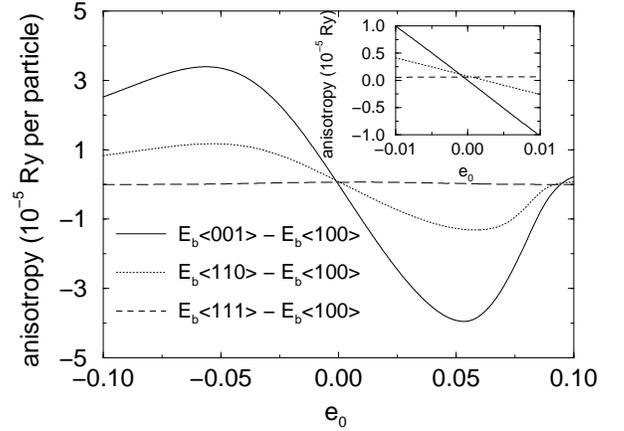}}
\vskip1cm \caption{Energy differences among $\langle001\rangle$,
$\langle100\rangle$, $\langle110\rangle$, and $\langle111\rangle$ magnetization
orientations vs. in-plane strain $e_0$ at $h=0.01$~Ry and $p = 3.5$~nm$^{-3}$.
For compressive strains ($e_0<0$) the systems has an easy
magnetic plane perpendicular to the growth direction. For  tensile strains
($e_0>0$) the anisotropy is easy-axis with the preferred magnetization
orientation along the growth direction. The anisotropy changes sign at 
large  tensile strain. } \label{anisostrain}
\end{figure}

\begin{figure}
\center
\epsfxsize 8.4cm {\epsffile{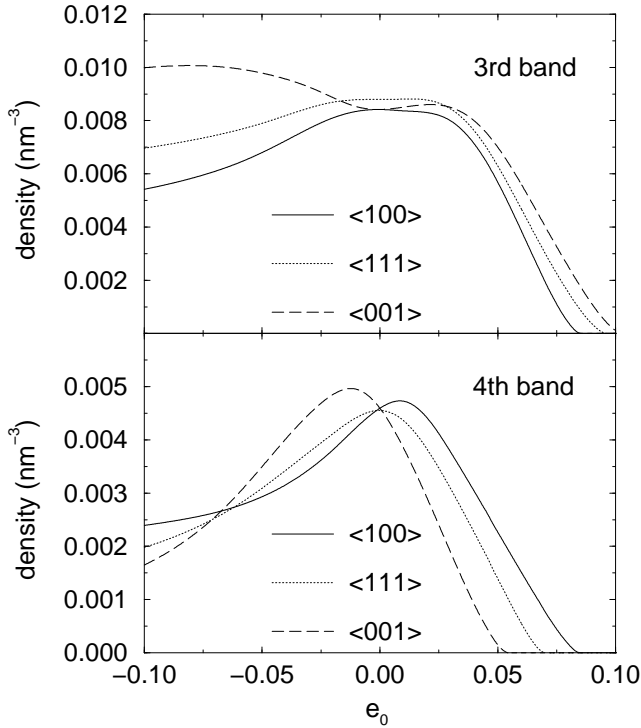}}
\vskip1cm \caption{3rd and 4th band densities for
$\langle100\rangle$, and $\langle111\rangle$, and
$\langle001\rangle$ magnetization
orientations vs. in-plane strain $e_0$ at $h=0.01$~Ry and $p = 3.5$~nm$^{-3}$.
Curves for the $\langle110\rangle$ magnetization orientation
(not shown here) are similar to 
those for the $\langle100\rangle$ orientation.}
\label{denstrain}
\end{figure}

In Fig.~\ref{anisostrain} we present mean-field theory predictions for the
strain-dependence of the anisotropy energy at $h=0.01{\rm Ry}$ and hole
density $p=0.35 {\rm nm}^{-3}$. According to our calculations, the easy axes in
the absence of strain are along the cube edges in this case. This calculation
is thus for a hole density approximately three times smaller than the Mn
density, as indicated by recent experiments.  The relevant value of $e_0$
depends on the substrate on which the epitaxial ${\rm Ga}_{.95}{\rm
Mn}_{.05}{\rm As}$ film is grown, as discussed in Section IV.  The most
important conclusion from Fig.~\ref{anisostrain} is that strains as small
as $1\%$ are sufficient to completely alter the magnetic anisotropy energy
landscape.  For example for (Ga,Mn)As on 
GaAs, $e_0=-.0028$ at $x=0.05$, the
anisotropy has a relatively strong uniaxial contribution even for this
relatively modest compressive strain, which favors in-plane moment
orientations, in agreement with experiment. A relatively small ($\sim$1 kJ 
m$^{-3}$) residual plane-anisotropy remains which favors $\langle110\rangle$
over $\langle100\rangle$.  For $x=0.05$ (Ga,Mn)As on a $x=0.15$ (In,Ga)As
buffer the strain is tensile, $e_0 =0.0077$, and we predict a substantial
uniaxial contribution to the anisotropy energy which favors growth direction
orientations, again in agreement with experiment.  For the
tensile case, the anisotropy energy changes more dramatically than for
compressive strains due to the depopulation of higher subbands, as
shown in Fig.~\ref{denstrain}. At large tensile strains, the sign of
the anisotropy changes
emphasizing the subtlety of these effects and the latitude which exists for
strain-engineering of magnetic properties.

\section{Discussion}

We first comment on the implications of the considerations described in this
paper for the interpretation of current experiments.  The hysteretic effects
which reflect magnetic anisotropy have been studied most extensively for the
highest $T_c$ samples currently available.  These mean-field-theory predictions
depend on three phenomenological parameters, $N_{Mn}$ which is sample dependent
but accurately known, $J_{pd}$ which should be nearly universal for a given
III-V host compound but is less accurately known, and the hole density $p$
which is sample dependent and not accurately known. Values of $J_{pd}$ and $p$
must be inferred from experiment, sometimes by comparison with theoretical
pictures which are not yet fully developed.  The reliability of $J_{pd}$ and
$p$ estimates is improving and presumably will continue to improve. It now
seems clear that the values of $J_{pd}$ and $p$ in current high $T_c$ samples
are such that several hole bands are partially occupied in the ferromagnetic
ground state.  In this case, we see from Fig.~\ref{Fig7} that our calculations
predict cube edge easy axes which include the growth direction. The anisotropy
energies are typically $~\sim 10^{-6} {\rm Ry} {\rm nm}^{-3} \sim 1 {\rm
kJ} {\rm m}^{-3}$.  Similar anisotropy energies are produced by strains as
small as $e_0 \sim 0.001$ and more typical strains produce larger anisotropy
energies.  An important conclusion from this work is that strain contributions
to the anisotropy will not normally be negligible.

We believe that the in-plane easy-axis observed in ${\rm Ga}_{1-x} {\rm Mn}_x
{\rm As}$ films grown on GaAs is a consequence of compressive strain in the magnetic
film which dominates the cubic band anisotropy energy. When ${\rm Ga}_{1-x}
{\rm Mn}_x {\rm As}$ is grown on (In,Ga)As, the lattice-matching strain is tensile,
reinforcing the  growth direction easy-axis anisotropy of strain free samples.
(As discussed earlier, the signs of both strain and cubic band contributions to
strain change when the light hole bands are depopulated.)  We note that in our
calculations, the cubic band anisotropy is almost always dominated by the
fourth cubic harmonic coefficient.  Given this, it follows that only the cube
edge easy axes, which includes the growth direction axis, and cube diagonal
axes, which are not in the film plane, are possible without strain.

Since the local moments are fully polarized in the ferromagnetic ground state,
it is easy to estimate the saturation moment $M_s \sim N_{Mn} g_L \mu_B J$,
leading to the relatively small numerical value $\mu_0 M_s \sim 0.05 {\rm T}$.
It follows that the growth direction orientation magnetostatic energy $\sim
\mu_0 M_s^2 \sim 0.1 {\rm kJ} {\rm m}^{-3}$, considerably smaller than the
magnetocrystalline anisotropy coefficient.  Even when several hole bands are
occupied with competing spin-orbit interactions, these materials have
relatively large magnetic hardness parameters.  Unlike the case of metallic
thin film ferromagnets which have much larger saturation moments, the
magnetostatic {\em shape anisotropy} plays a minor role in the dependence of
total energy on moment orientation.  The small saturation moment, will also
tend to lead to large domain sizes and square easy-axis direction hysterisis
loops, as seen in experiment.

Coercivities can be estimated\cite{coeyskomski} from the anisotropy fields
defined by
\begin{equation}
\mu_0 H_a \sim \mu_0 M_s \frac{ K}{\mu_0 M_s^2}.
\end{equation}
For hard magnetic materials, anisotropy fields are much larger than saturation
magnetizations.  The itinerant field places an upper bound on and is expected
to scale with the coercivity. Our calculations suggest that the coercivity in
ferromagnetic samples with a single partially occupied hole band will be
immensely larger than the coercivity of current samples.  Such samples could be
fabricated, for example, by adding donors such as Si to current samples,
further compensating the Mn acceptors.  According to mean-field-theory, this
modification in the sample preparation procedure will lower the ferromagnetic
critical temperature, and at the same time, increase the anisotropy energy.

Finally we conclude with a few words of caution.  This theory of magnetic
anisotropy has three principle limitations: i) it is based on a mean-field
theory description of the exchange interaction between localized spins and
valence band holes, ii) it neglects hole-hole interactions, and iii) it doesn't
account for disorder scattering of the itinerant holes.  Confronting these
weaknesses would in each case considerably complicate the theory and we feel it
is appropriate to seek progress by comparing the present relatively simple
theory with experiment. Nevertheless it is worthwhile to speculate on where and
how the theory may be expected to fail.

Mean-field theory should be reliable when the range of the hole mediated
interaction between localized spins is long {\em and} the spin-stiffness
parameter that characterizes the energy of long-wavelength spin-fluctuations is
sufficiently large.  Considerations of this type suggest\cite{juergen} that
mean-field theory will fail at high temperatures unless the ratio of the hole
density to the localized spin density is small and the field $h$ is not too
large compared to the Fermi energy.  Since $p$ is typically smaller than
$N_{Mn}$ because of anti-site defects in low-temperature MBE growth samples,
mean-field theory is likely to be reasonable at least at $T=0$ in many
(III,Mn)V ferromagnets.

Hole-hole interactions will clearly tend to favor the ferromagnetic state by
countering the band-energy cost of the spin polarization. Because of strong
spin-orbit coupling in the valence band, estimates based on many-body
calculations for electron gas systems may be of little use in estimating the
importance of this effect more quantitatively.  Work is currently in progress
which should shed more light on this matter\cite{tomas}. Nevertheless, it seems
likely that these interactions will not have an overriding importance, at least
in large hole density samples.

Finally we come to disorder.  It is clear from experiment, that disorder does
not have a qualitative impact on free-carrier mediated ferromagnetism even when
those free carriers have been localized by a random disorder potential. It
seems likely that disorder will destroy the ferromagnetic state, only when the
localization length becomes comparable to the distance between localized spins.
On the other hand, since elastic disorder scattering will mix band states with
different orientations on the Fermi surface, it also seems clear that a
reduction in magnetic anisotropy energy must result.  Indeed the coercivities
that follow from our anisotropy energy results appear to be larger than what is
observed.  As far as we are aware, no theory of this effect exists at present.

\acknowledgements

We gratefully acknowledge helpful interactions with W.A. Atkinson, T. Dietl, J.
Furdyna, J.A. Gaj, J.  K\" onig, B.-H. Lee, E. Miranda, Hsiu-Hau Lin, and Hideo
Ohno. Work at the University of Oklahoma was supported by the NSF under grant
No. EPS-9720651 and a grant from the Oklahoma State Regents for Higher
Education.  Work at Indiana University was performed under NSF grants
DMR-9714055 and DGE-9902579.  Work at the Institute of Physics ASCR was
supported by the Grant Agency of the Czech Republic under grant 202/98/0085.
AHM gratefully acknowledges the hospitality of UNICAMP where project work was
initiated.

\section*{Appendix}
In the literature, different representations are used for the four band and six
band model Kohn-Luttinger Hamiltonians.  In the interest of completeness and
clarity, this appendix specifies the expressions on which our detailed
calculations are based.  Detailed derivations of ${\bf k}\cdot{\bf p}$
perturbation theory for cubic semiconductors can be found
elsewhere.\cite{valencebands}

The $k=0$ states at the top valence band have $p$-like character and can be
represented by the $l=1$ orbital angular momentum eigenstates $|m_l\rangle$. In
the coordinate representation we can write
\begin{eqnarray}
\langle{\bf r}|m_l=1\rangle&=&-\frac{1}{\sqrt{2}}f(r)(x+iy) \nonumber \\
\langle{\bf r}|m_l=0\rangle&=&f(r)z \nonumber \\ \langle{\bf
r}|m_l=-1\rangle&=&\frac{1}{\sqrt{2}}f(r)(x-iy). \label{oam}
\end{eqnarray}
The Kohn-Luttinger Hamiltonian for systems with no spin-orbit coupling, ${\cal
H}_L$, is written in the representation of the following combinations of
$|m_l\rangle$
\begin{eqnarray}
|X\rangle&=&\frac{1}{\sqrt{2}}\big(|m_l=-1\rangle-|m_l=1\rangle\big) \nonumber
\\ |Y\rangle&=&\frac{i}{\sqrt{2}}\big(|m_l=-1\rangle+|m_l=1\rangle\big)
\nonumber \\ |Z\rangle&=&|m_l=0\rangle\; . \label{xyz}
\end{eqnarray}
It reads
\begin{equation}
\hspace*{-.5cm}{\cal H}_L = \left(\begin{array}{ccc} Ak_x^2+B(k_y^2+k_z^2) &
Ck_xk_y & Ck_xk_z \\ Ck_yk_x & Ak_y^2+B(k_x^2+k_z^2) & Ck_yk_z \\ Ck_zk_x &
Ck_zk_y & Ak_z^2+B(k_x^2+k_y^2)  \\
\end{array}\right),
\label{noso}
\end{equation}
where
\begin{eqnarray}
A&=&-\frac{\hbar^2}{2m}(\gamma_1+4\gamma_2), \nonumber \\
B&=&-\frac{\hbar^2}{2m}(\gamma_1-2\gamma_2), \nonumber \\
C&=&-\frac{3\hbar^2}{m}\gamma_3, \label{abc}
\end{eqnarray}
$m$ is the bare electron mass, and $\gamma_1$, $\gamma_2$, and $\gamma_3$ are
the phenomenological Luttinger parameters. To include spin-orbit coupling we
use the basis formed by total angular momentum eigenstates $|j,m_j\rangle$:
\begin{eqnarray}
|1\rangle&\equiv&|j=3/2,m_j=3/2\rangle \nonumber \\
|2\rangle&\equiv&|j=3/2,m_j=-1/2\rangle \nonumber \\
|3\rangle&\equiv&|j=3/2,m_j=1/2\rangle \nonumber \\
|4\rangle&\equiv&|j=3/2,m_j=-3/2\rangle \nonumber \\
|5\rangle&\equiv&|j=1/2,m_j=1/2\rangle \nonumber \\
|6\rangle&\equiv&|j=1/2,m_j=-1/2\rangle \label{123456}
\end{eqnarray}
The basis (\ref{123456}) is related to the orbital angular momentum
($m_l=1,0,-1$) and spin ($\sigma=\uparrow,\downarrow$) eigenstates by
\begin{eqnarray}
|1\rangle&=&|m_l=1,\uparrow\rangle \nonumber \\
|2\rangle&=&\frac{1}{\sqrt{3}}\, |m_l=-1,\uparrow\rangle+\sqrt{\frac23}\,
|m_l=0,\downarrow\rangle \nonumber \\ |3\rangle&=&\frac{1}{\sqrt{3}}\,
|m_l=1,\downarrow\rangle+\sqrt{\frac23}\, |m_l=0,\uparrow\rangle \nonumber \\
|4\rangle&=&|m_l=-1,\downarrow\rangle \nonumber \\
|5\rangle&=&-\frac{1}{\sqrt{3}}\, |m_l=0,\uparrow\rangle+\sqrt{\frac23}\,
|m_l=1,\downarrow\rangle \nonumber \\ |6\rangle&=&\frac{1}{\sqrt{3}}\,
|m_l=0,\downarrow\rangle-\sqrt{\frac23}\, |m_l=-1,\uparrow\rangle \label{jls}
\end{eqnarray}
or
\begin{eqnarray}
|1\rangle&=&-\frac{1}{\sqrt{2}}\big(|X,\uparrow\rangle
+i|Y,\uparrow\rangle\big) \nonumber \\
|2\rangle&=&\frac{1}{\sqrt{6}}\big(|X,\uparrow\rangle
-i|Y,\uparrow\rangle\big)+\sqrt{\frac23}\, |Z,\downarrow\rangle \nonumber \\
|3\rangle&=&-\frac{1}{\sqrt{6}}\big(|X,\uparrow\rangle
+i|Y,\uparrow\rangle\big)+\sqrt{\frac23}\, |Z,\uparrow\rangle \nonumber \\
|4\rangle&=&\frac{1}{\sqrt{2}}\big(|X,\downarrow\rangle
-i|Y,\downarrow\rangle\big) \nonumber \\
|5\rangle&=&-\frac{1}{\sqrt{3}}\big(|X,\downarrow\rangle
+i|Y,\downarrow\rangle\big)-\frac{1}{\sqrt{3}}\, |Z,\uparrow\rangle \nonumber
\\ |6\rangle&=&-\frac{1}{\sqrt{3}}\big(|X,\uparrow\rangle
-i|Y,\uparrow\rangle\big)+\frac{1}{\sqrt{3}}\, |Z,\downarrow\rangle \label{jrs}
\end{eqnarray}
The six band model Kohn-Luttinger Hamiltonian, $H_L$, in the representation of
vectors (\ref{123456}) is
\begin{equation}
\hspace*{0cm} H_L = \left(\begin{array}{cccccc} {\cal H}_{hh} & -c & -b &
\multicolumn{1}{c|}{0} & \frac{b}{\sqrt{2}} & c\sqrt{2}\\ -c^* & {\cal H}_{lh}
& 0 & \multicolumn{1}{c|}{b} & -\frac{b^*\sqrt{3}}{\sqrt{2}} & -d\\ -b^* & 0 &
{\cal H}_{lh} & \multicolumn{1}{c|}{-c} &   d & -\frac{b\sqrt{3}}{\sqrt{2}} \\
0 & b^* & -c^* & \multicolumn{1}{c|}{{\cal H}_{hh}} &  -c^*\sqrt{2} &
\frac{b^*}{\sqrt{2}}\\ \cline{1-4} \frac{b^*}{\sqrt{2}} &
-\frac{b\sqrt{3}}{\sqrt{2}} & d^* & -c\sqrt{2} & {\cal H}_{so} & 0\\
c^*\sqrt{2} & -d^* & -\frac{b^*\sqrt{3}}{\sqrt{2}} & \frac{b}{\sqrt{2}} & 0 &
{\cal H}_{so}\\
\end{array}\right)
\label{hl}
\end{equation}
In the matrix (\ref{hl}) we highlighted the four band model Hamiltonian. The
Kohn-Luttinger eigenenergies are measured down from the top of the valence
band, i.e., they are hole energies and we use the following notation:
\begin{eqnarray}
{\cal H}_{hh} &=& \frac{\hbar^2}{2m}\big[(\gamma_1 + \gamma_2)(k_x^2+k_y^2) +
(\gamma_1 - 2\gamma_2)k_z^2 \nonumber \\ {\cal H}_{lh} &=&
\frac{\hbar^2}{2m}\big[(\gamma_1 - \gamma_2)(k_x^2+k_y^2) + (\gamma_1 +
2\gamma_2)k_z^2 \nonumber \\ {\cal H}_{so} &=&
\frac{\hbar^2}{2m}\gamma_1(k_x^2+k_y^2+k_z^2) + \Delta_{so} \nonumber \\ b &=&
\frac{\sqrt{3}\hbar^2}{m} \gamma_3 k_z (k_x - i k_y) \nonumber \\ c &=&
\frac{\sqrt{3}\hbar^2}{2m}\big[\gamma_2(k_x^2 - k_y^2) - 2i\gamma_3 k_x
k_y\big] \nonumber \\ d &=&
-\frac{\sqrt{2}\hbar^2}{2m}\gamma_2\big[2k_z^2-(k_x^2 + k_y^2 )\big]\; .
\label{lutpar}
\end{eqnarray}

The four band Kohn-Luttinger Hamiltonian can be diagonalized analytically and
yields a pair of Kramers doublets with eigenenergies
\begin{eqnarray}
\varepsilon_{{\bf k}} = \frac{H_{hh} + H_{lh}}{2} \mp \sqrt{\frac{1}{4}(H_{hh}
- H_{lh})^2 + |b|^2 + |c|^2}. \nonumber
\end{eqnarray}
In the spherical approximation\cite{spherical} ($\gamma_2,\gamma_3 \to \bar
\gamma \equiv 0.6 \gamma_2 + 0.4 \gamma_3$), the top of the valence band
consists of two doubly degenerate parabolic bands with effective masses
$m_h=m/(\gamma_1-2 \bar \gamma) \sim 0.498 m$ and $m_l = m/(\gamma_1+\ 2 \bar
\gamma) \sim 0.086 m$.  (An approximation in which the light hole bands, which
have a much smaller density of states, are ignored is adequate for some
purposes.)

The four band and six band representations for the hole spin-operator
components read
\begin{eqnarray}
s_x &=& \left(\begin{array}{cccccc} 0 & 0 & \frac{1}{2\sqrt{3}} &
\multicolumn{1}{c|}{0} & \frac{1}{\sqrt{6}} & 0\\ 0 & 0 & \frac13 &
\multicolumn{1}{c|}{\frac{1}{2\sqrt{3}}} & -\frac{1}{3\sqrt{2}} & 0 \\
\frac{1}{2\sqrt{3}} & \frac13 & 0 & \multicolumn{1}{c|}{0} & 0 &
\frac{1}{3\sqrt{2}} \\ 0 & \frac{1}{2\sqrt{3}} & 0 & \multicolumn{1}{c|}{0} & 0
& -\frac{1}{\sqrt{6}} \\ \cline{1-4} \frac{1}{\sqrt{6}} & -\frac{1}{3\sqrt{2}}
& 0 & 0 & 0 & -\frac16\\ 0 & 0 & \frac{1}{3\sqrt{2}} & -\frac{1}{\sqrt{6}} &
-\frac16 & 0
\end{array}\right) \nonumber \\ & &\nonumber \\  & &\nonumber \\
s_y &=& i\left(\begin{array}{cccccc} 0 & 0 & -\frac{1}{2\sqrt{3}} &
\multicolumn{1}{c|}{0} & -\frac{1}{\sqrt{6}} & 0\\ 0 & 0 & \frac13 &
\multicolumn{1}{c|}{-\frac{1}{2\sqrt{3}}} & -\frac{1}{3\sqrt{2}} & 0 \\
\frac{1}{2\sqrt{3}} & -\frac13 & 0 & \multicolumn{1}{c|}{0} & 0 &
-\frac{1}{3\sqrt{2}} \\ 0 & \frac{1}{2\sqrt{3}} & 0 & \multicolumn{1}{c|}{0} &
0 & -\frac{1}{\sqrt{6}} \\ \cline{1-4} \frac{1}{\sqrt{6}} & \frac{1}{3\sqrt{2}}
& 0 & 0 & 0 & \frac16\\ 0 & 0 & \frac{1}{3\sqrt{2}} & \frac{1}{\sqrt{6}} &
-\frac16 & 0
\end{array}\right) \nonumber \\ & &\nonumber \\ & &\nonumber \\
s_z &=& \left(\begin{array}{cccccc} \frac12 & 0 & 0 & \multicolumn{1}{c|}{0} &
0 & 0\\ 0 & -\frac16 & 0 & \multicolumn{1}{c|}{0} & 0 & -\frac{\sqrt{2}}{3} \\
0 & 0 & \frac16 & \multicolumn{1}{c|}{0} & -\frac{\sqrt{2}}{3} & 0 \\ 0 & 0 & 0
& \multicolumn{1}{c|}{-\frac12} & 0 & 0 \\ \cline{1-4} 0 & 0 &
-\frac{\sqrt{2}}{3} & 0 &  -\frac16 & 0\\ 0 &  -\frac{\sqrt{2}}{3} & 0 & 0 & 0
& \frac16
\end{array}\right)
\label{spin}
\end{eqnarray}

%%%%%%%%%%%%%%%%%%%%%%%%%%%%%%%%%%%%%%%%%%%%%%%%%%%%%%%%%%%%%%%%%%%%%%%%%%%%%%

\begin{table}
\begin{center}
\begin{tabular}{cccccc}
Host & $\bar \gamma<100>$ & $\bar \gamma<110>$ & $\bar \gamma<111>$ & $
\gamma^{ca}_{1}$ & $ \gamma^{ca}_{2}$ \\ \hline GaAs & 5.965 & 5.088 & 4.639 &
-3.509 & -4.24 \\ InAs & 13.207 & 10.854 & 9.705 & -9.412 & -9.84
\end{tabular}
\end{center}
\caption{\protect High symmetry direction moment-orientation dependent average
Luttinger parameters and their cubic harmonic expansions for GaAs and InAs
based ${\rm III}_{1-x}{\rm Mn}_{x}{\rm V}$ ferromagnetic semiconductors. These
parameters specify the $T=0$  magnetic anisotropy energy in the limit of large
spin-orbit splitting and large exchange coupling parameter ($J_{pd}$) or small
hole density $p$. } \label{anisotropy}
\end{table}

%%%%%%%%%%%%%%%%%%%%%%%%%%%%%%%%%%%%%%%%%%%%%%%%%%%%%%%%%%%%%%%%%%%%%%%%%%%%%%
%\end{multicols}
%%%%%%%%%%%%%%%%%%%%%%%%%%%%%%%%%%%%%%%%%%%%%%%%%%%%%%%%%%%%%%%%%%%%%
\end{document}